\documentclass[10pt,preprintnumbers,showpacs,amsmath,amssymb,floatfix,twocolumn]{revtex4}   
\usepackage{graphicx}
\usepackage{dcolumn}
\usepackage{bm}
\usepackage{longtable}
\usepackage{graphics}
\usepackage{amssymb}
\usepackage{amsmath}
\usepackage{xspace}
\usepackage{epsfig}
\begin{document}
\title{5,6-dihydroxyindole-2-carboxylic acid (DHICA): a
First Principles Density-Functional Study}
\author{B. J. Powell}

\email{powell@physics.uq.edu.au}

\affiliation{Department of Physics, University of Queensland,
Brisbane, Queensland 4072, Australia}


\begin{abstract}
We report first principles density functional calculations for
5,6-dihydroxyindole-2-carboxylic acid (DHICA) and several reduced
forms. DHICA and 5,6-dihydroxyindole (DHI) are believed to be the
basic building blocks of the eumelanins. Our results show that
carboxylation has a significant effect on the physical properties
of the molecules. In particular, the relative stabilities and the
HOMO-LUMO gaps (calculated with the $\Delta$SCF method) of the
various redox forms are strongly affected. We predict that, in
contrast to DHI, the density of unpaired electrons, and hence the
ESR signal, in DHICA is negligibly small.
\end{abstract}

\maketitle

\section{Introduction}

The melanins are an important class of pigmentary macromolecule
found throughout nature from fungi to man \cite{Prota}. Two types
of melanin are found in the skin, hair, eyes and nervous systems
of humans: pheomelanin and eumelanin. Pheomelanin is associated
with ginger hair and is a cysteinyl-dopa derivative. Eumelanin,
which we consider here, is found in much higher levels in black
hair and is believed to be a macromolecule of
5,6-dihydroxyindole-2-carboxylic acid (DHICA) and
5,6-dihydroxyindole (DHI) \cite{Prota,Ito}. Perhaps the most
surprising property of melanins is that, whereas most biomolecules
show only well defined absorption peaks around 280 nm (4.5 eV) and
little absorption at lower energies \cite{vanHolde}, the melanins
show broad band monotonic absorption in the range 1.5 to 5 eV
\cite{Wolbarsht}. Melanins are also efficient free radical
scavengers and antioxidants \cite{Prota}. Thus melanins are
ideally suited for their best understood biological function as
the primary photoprotectant in our skin and eyes. Conversely, both
pheomelanin and eumelanin are implicated in the development of
melanoma skin cancer \cite{Hill_mel}. Melanins also display
semiconductor like behaviour \cite{McGinness,Crippa,Jastrzebska}
that suggests possible uses in applications such as bio-sensors
and bio-mimetic photovoltaics
\cite{Paulschapter,Meredith&Riesz03}. However, it is not clear
that the conductivity reported in any of these experimental
studies is electronic in nature \cite{Paulschapter,ICSM04}.

Surprisingly, given the obvious importance of the melanins and
several decades of work, little is known about the general
structure-property-function relationships that control their
behaviour. This is, at least in part, due to the difficulties in
studying the melanins; they are chemically and photochemically
stable, and are virtually insoluble in most common solvents. For
example, major questions still remain concerning their basic
structural unit \cite{Zajac}. Two opposing schools of thought
exist: i) that eumelanins are composed of highly cross-linked
extended hetero-polymers based upon the Raper-Mason scheme
\cite{Prota}, and ii) that eumelanins are actually composed of
much smaller oligomers condensed into 4 or 5 oligomer
nano-aggregates \cite{Clancy}.
%
A clear idea of the basic structural unit is critical to
developing a consistent model for condensed phase charge transport
in such disordered organic systems. It is also important in the
context of \lq\lq molecularly engineering" melanins to have the
ability to create or enhance functionality in high technology
applications \cite{Paulschapter,Meredith&Riesz03}.

Recently the role of disorder in determining the physical
properties of the melanins has been emphasised
\cite{mel_mon,Paulschapter,Bochenek}. In particular it has been
proposed that the broad band monotonic optical absorption of
eumelanin arises from the existence of a large number of different
monomers and hence a wide variety of macromolecules. The idea is,
basically, that, although each of these molecules has a distinct
optical absorption that is not atypical for bio-macromolecule,
summing over a large range of absorption frequencies in a
biological environment (in which both dipole fluctuations
\cite{Joel} and thermal effects broaden the spectrum) gives rise
to the observed broad band monotonic absorption spectrum.

\begin{figure}
    \centering
    \epsfig{figure=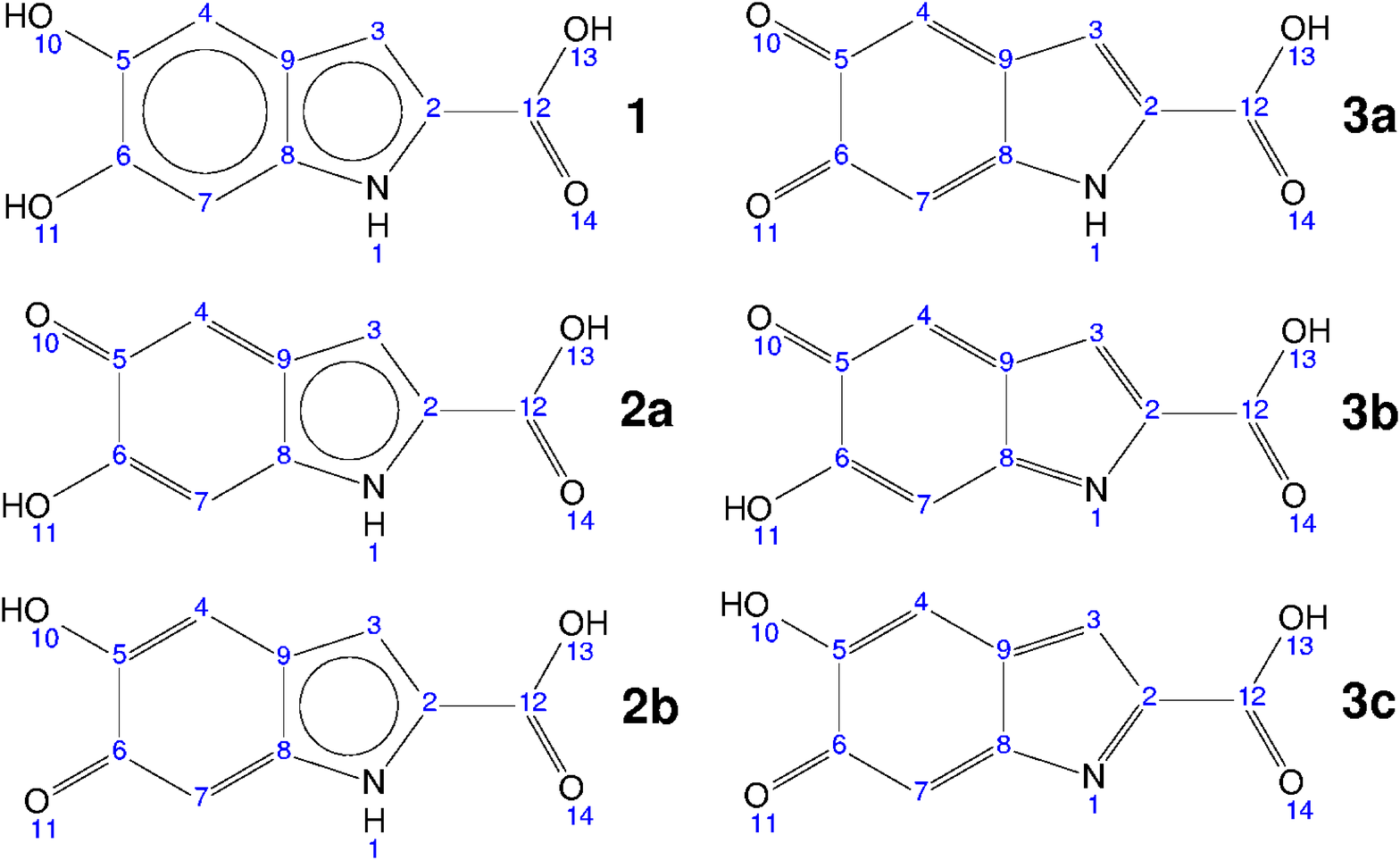, width=8.5cm, angle=0}
    \caption{Schematic representation of
    5,6-dihydroxyindole-2-carboxylic acid (DHICA) {\bf 1}
    and the reduced forms which we also consider here.
    The numbers correspond to those used in table \ref{tab:bond_lengths}.
    Note that {\bf 2a} and {\bf 2b} are radicals in the neutral state.} \label{fig:struct}
\end{figure}

Despite the widespread agreement that natural eumelanin contains
both DHI and DHICA, previous quantum chemical studies
\cite{mel_mon,Pullman,GC1,GC2,GC3,Bolivar-Marinez,Bochenek,Stark,Ilichev}
have only considered DHI. This is in part due to the greater
chemical complexity of DHICA, but also in part due to assumption
that the carboxylation of DHI does not play a significantly role
in determining the properties eumelanin. However, given the
potential importance of the role of disorder in the physical
properties of eumelanin \cite{mel_mon,Paulschapter,Bochenek} a
careful study of the similarities and differences between DHI and
DHICA is clearly required. Our study is also motivated by recent
work on synthetic DHICA \cite{Tran} and the need for high quality
first principles calculations to compare with future experiments
on synthetic DHICA. In this paper we present, to our knowledge,
the first quantum chemical study of DHICA. We use density
functional theory (DFT) to calculate the physical and electronic
structure of DHICA and several reduced forms, each of which is
drawn schematically in figure \ref{fig:struct}. Our results show
that there are significant differences between the physical
properties of DHI and DHICA. Specifically we show that the
relative stability of the reduced forms is significantly altered
by carboxylation as is the HOMO-LUMO gap [the energy difference
between the highest occupied molecular orbital (HOMO) and the
lowest unoccupied molecular orbital (LUMO)].

\section{Calculation details}

The chemical and electronic structures were found from first
principles DFT calculations. We have performed our calculations
using the Naval Research Laboratory Molecular Orbital Library
(NRLMOL)
\cite{NRLMOL1,NRLMOL2,NRLMOL3,NRLMOL4,NRLMOL5,NRLMOL6,NRLMOL7}.
NRLMOL performs massively parallel electronic structure
calculation using gaussian orbital methods. Throughout we have
used the Perdew, Burke and Ernzerhof (PBE) \cite{PBE} exchange
correlation functional, which is a generalised gradient
approximation (GGA)
containing no parameters. 
For each molecule we have fully relaxed the geometry with no
symmetry constraints.

\section{Results and Discussion}\label{sect:results}


We report the calculated bond lengths and bond angles of DHICA in
table \ref{tab:bond_lengths}. For comparison we also report the
same quantities for DHI from calculations \cite{mel_mon} using the
same methods (it should be noted that the results for DHI are in
good agreement with other quantum chemical calculations
\cite{Bolivar-Marinez,Stark,Ilichev}). Unsurprisingly there is
little difference between the physical structure of DHI and DHICA.

\begin{table}
\caption{\label{tab:bond_lengths} The calculated bond lengths (in
\AA) and bond angles (in degrees) of DHICA. The atom numbers
correspond to those shown in figure \ref{fig:struct}. Note that
there is no experimental data or previous theoretical results to
compare with. We therefore contrast our results with previously
published results for DHI \cite{mel_mon}.}
\begin{ruledtabular}
\begin{tabular}{lcc}
 & DHICA  & DHI  \\
 \hline\vspace*{-9pt} \\
N1-C2 & 1.387 &  1.390 \\
C2-C3 & 1.388 &  1.365 \\
C3-C9 & 1.421 &  1.443 \\
C9-C4 & 1.414 & 1.366 \\
C4-C5 & 1.380 & 1.458 \\
C5-C6 & 1.428 & 1.586 \\
C5-O10 & 1.384 &  1.226 \\
C6-O11 & 1.365 &  1.226 \\
C6-C7 & 1.387 &  1.457 \\
C7-C8 & 1.403 &  1.361 \\
C8-C9 & 1.432 & 1.483 \\
C8-N1 & 1.372 &  1.381 \\
C2-C12 & 1.453 &  - \\
C12-O13 & 1.368 &  - \\
C12-O14 & 1.225 &  - \\
N1-C2-C3 & 109.1 & 110.8 \\
C2-C3-C9 & 107.0 & 107.2 \\
C3-C9-C4 & 133.5 & 132.7 \\
C3-C9-C8 & 107.2 & 106.2 \\
C9-C4-C5 & 118.3 & 119.7 \\
C4-C5-C6 & 121.7 &  118.3 \\
C4-C5-O10 & 124.8 & 122.7 \\
O10-C5-C6 & 113.6 & 118.8 \\
C5-C6-O11 & 118.9 & 119.1 \\
C5-C6-C7 & 121.2 & 117.9 \\
O11-C6-C7 & 119.8 & 122.8 \\
C6-C7-C8 & 117.2 & 118.3 \\
C7-C8-C9 & 122.3 & 124.6 \\
C8-C9-C4 & 119.3 & 121.0 \\
N1-C8-C9 & 107.3 & 106.0 \\
C7-C8-N1 & 130.5 & 129.3 \\
C8-N1-C2 & 109.4 & 109.6 \\
N1-C2-C12 & 118.9 & - \\
C3-C2-C12 & 132.0 & - \\
C2-C12-O13 & 112.9 & - \\
C2-C12-O14 & 124.2 & - \\
O13-C12-O14 & 112.8 & - \\
\end{tabular}
\end{ruledtabular}
\end{table}


In table \ref{tab:energies} we report the calculated energies of
each of the molecules considered here in there neutral, -1 and -2
charge states. We also report the relative concentration at 300~K
in table \ref{tab:abundances}. It can be seen that in the neutral
state DHICA is considerably more stable than any of the reduced
forms. Whereas in the -1 state {\bf 2a} and {\bf 2b} are most
stable. This result has important implications for electron spin
resonance (ESR) experiments. ESR experiments are sensitive to
unpaired electrons which, these calculations indicate, are not
found in DHICA (this is not the case for DHI
\cite{Ilichev,mel_mon}). Given the recent synthesis of DHICA, ESR
experiments on pure DHICA samples would provide a direct check of
these calculations. These results suggest that samples of DHICA
have significantly less chemical disorder than pure DHI or natural
eumelanin. We can therefore speculate that the optical absorption
of macromolecules formed from synthetic DHICA may not be a smooth
as that of natural eumelanin. Note that for all of the molecules
considered here the -1 charge state is more stable than the
neutral state (as is the case for the related DHI molecules
\cite{Bochenek,mel_mon}), this is clearly related to free radical
scavenging properties of eumelanin.


\begin{table}
\caption{\label{tab:energies} The total energy (in eV) of DHICA
and the reduced forms shown in figure \ref{fig:struct} in the
neutral, -1 and -2 charge states. The energies are quoted relative
to the energy of neutral DHICA. In each case we fully relaxed the
geometries as shown in figure \ref{fig:density}.}
\begin{ruledtabular}
\begin{tabular}{lccc}
& Neutral
& -1 & -2 \\
 \hline\vspace*{-9pt} \\
{\bf 1} & 0 & -0.299 & +3.245 \\
{\bf 2a} & +1.165 & -1.478 & +1.887 \\
{\bf 2b} & +1.152 & -1.613 & +1.813 \\
{\bf 3a} & +2.392 & -0.223 & +1.814 \\
{\bf 3b} & +2.693 & -0.307 & +1.273 \\
{\bf 3c} & +2.583 & -0.321 & +1.276 \\
\end{tabular}
\end{ruledtabular}
\end{table}

\begin{table}
\caption{\label{tab:abundances} The relative abundances of DHICA
and the reduced forms shown in figure \ref{fig:struct} in the
neutral, -1 and -2 charge states at 300~K. We have assumed a
Boltzmann distribution of each charge state separately with an
infinite reservoir of H$_2$. The relevant energies are stated in
table \ref{tab:energies}.}
\begin{ruledtabular}
\begin{tabular}{lccc}
& Neutral
& -1 & -2 \\
 \hline\vspace*{-9pt} \\
{\bf 1} & 1 & $<10^{-22}$ & $<10^{-33}$ \\
{\bf 2a} & $<10^{-19}$ & $5.5\times 10^{-3}$ & $<10^{-10}$ \\
{\bf 2b} & $<10^{-19}$ & 1 & $<10^{-9}$ \\
{\bf 3a} & $<10^{-40}$ & $<10^{-23}$ & $<10^{-5}$ \\
{\bf 3b} & $<10^{-45}$ & $<10^{-21}$ & 1 \\
{\bf 3c} & $<10^{-44}$ & $<10^{-21}$ & 0.88 \\
\end{tabular}
\end{ruledtabular}
\end{table}

\begin{table}
\caption{\label{tab:HOMO-LUMO} The HOMO-LUMO (highest occupied
molecular orbital-lowest unoccupied molecular orbital) gap in eV.
The HOMO-LUMO gap has been calculated from both the $\Delta$SCF
method and from a simple interpretation of the Kohn--Sham
eigenvalues. We have previously shown \cite{mel_mon,Paulschapter}
that for DHI and related molecules the value of the HOMO-LUMO gap
calculated by $\Delta$SCF method is in good agreement with the
results of time dependent density functional theory (TDDFT)
calculations for the same molecules \cite{Ilichev}. The figures in
brackets show the results of equivalent calculations for DHI
\cite{mel_mon}. It can be seen that there is a significant change
in the HOMO-LUMO gap for {\bf 1}, which is essentially the only
stable monomer in the neutral state (c.f. tables
\ref{tab:energies} and \ref{tab:abundances}).}
\begin{ruledtabular}
\begin{tabular}{lcc}
 & \begin{tabular}{c} Simple interpretation of the \\Kohn--Sham eigenvalues\end{tabular} & $\Delta$SCF  \\
 \hline\vspace*{-9pt} \\
{\bf 1}  & 2.85 (3.48) & 3.04 (3.61) \\
{\bf 2a} & 2.24 (-) & 2.67 (-) \\
{\bf 2b} & 2.37 (-) & 2.64 (-) \\
{\bf 3a} & 0.87 (1.07) & 1.96 (2.02) \\
{\bf 3b} & 0.79 (0.80) & 1.10 (1.12) \\
{\bf 3c} & 0.89 (-) & 1.25 (-) \\
\end{tabular}
\end{ruledtabular}
\end{table}

\begin{figure}
    \centering
    \epsfig{figure=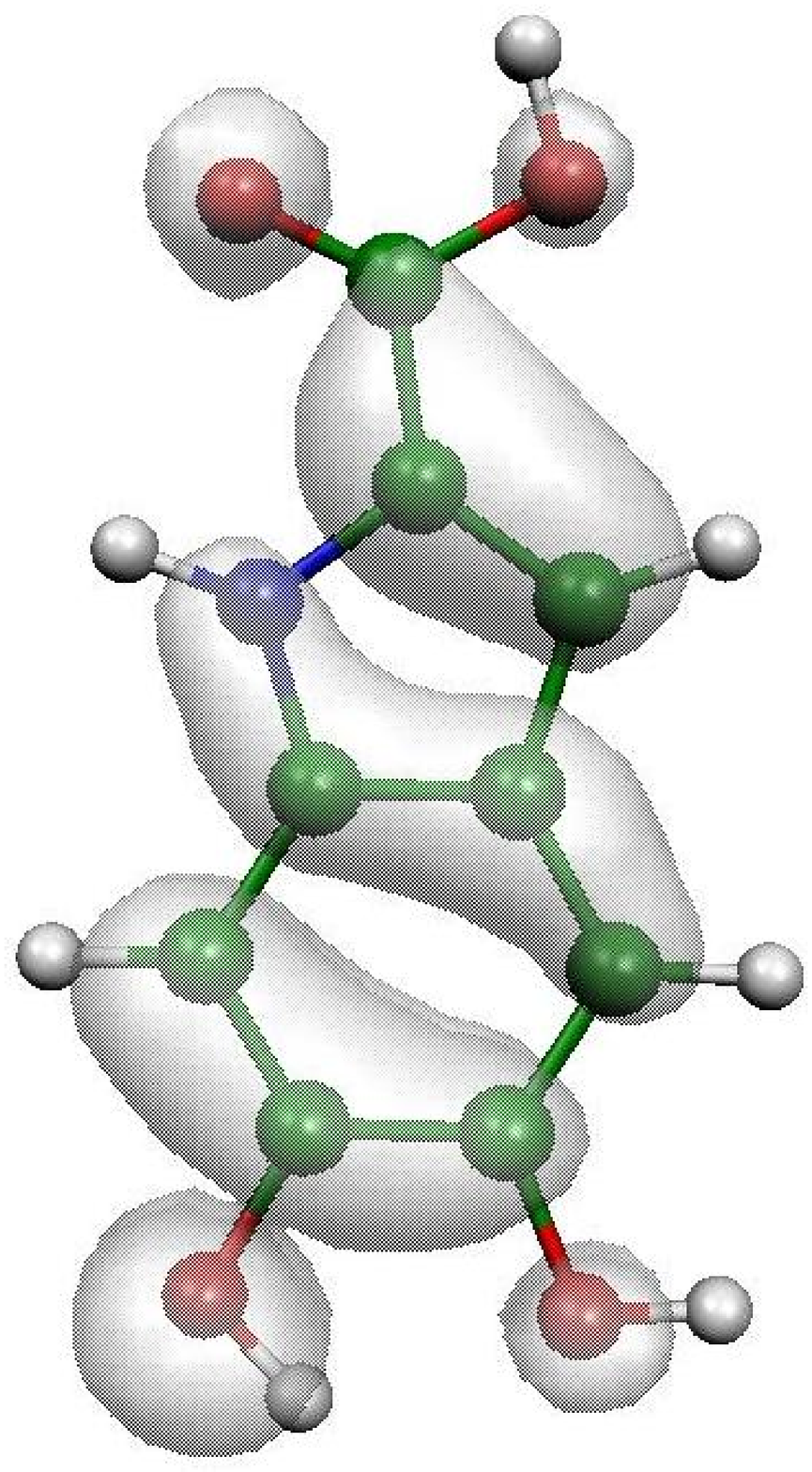, width=2.3cm, angle=90}
    \epsfig{figure=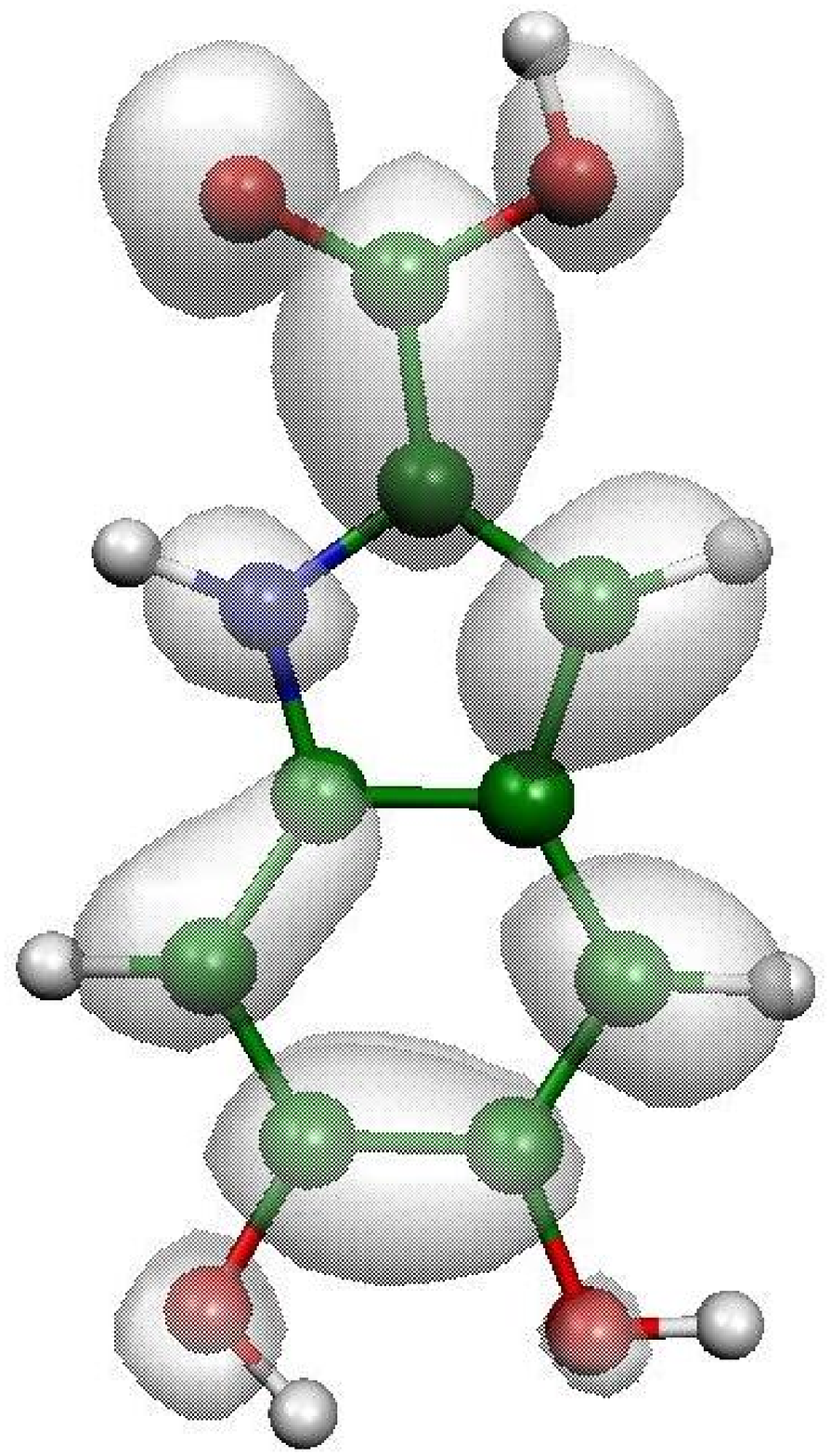, width=2.3cm, angle=90}
    \epsfig{figure=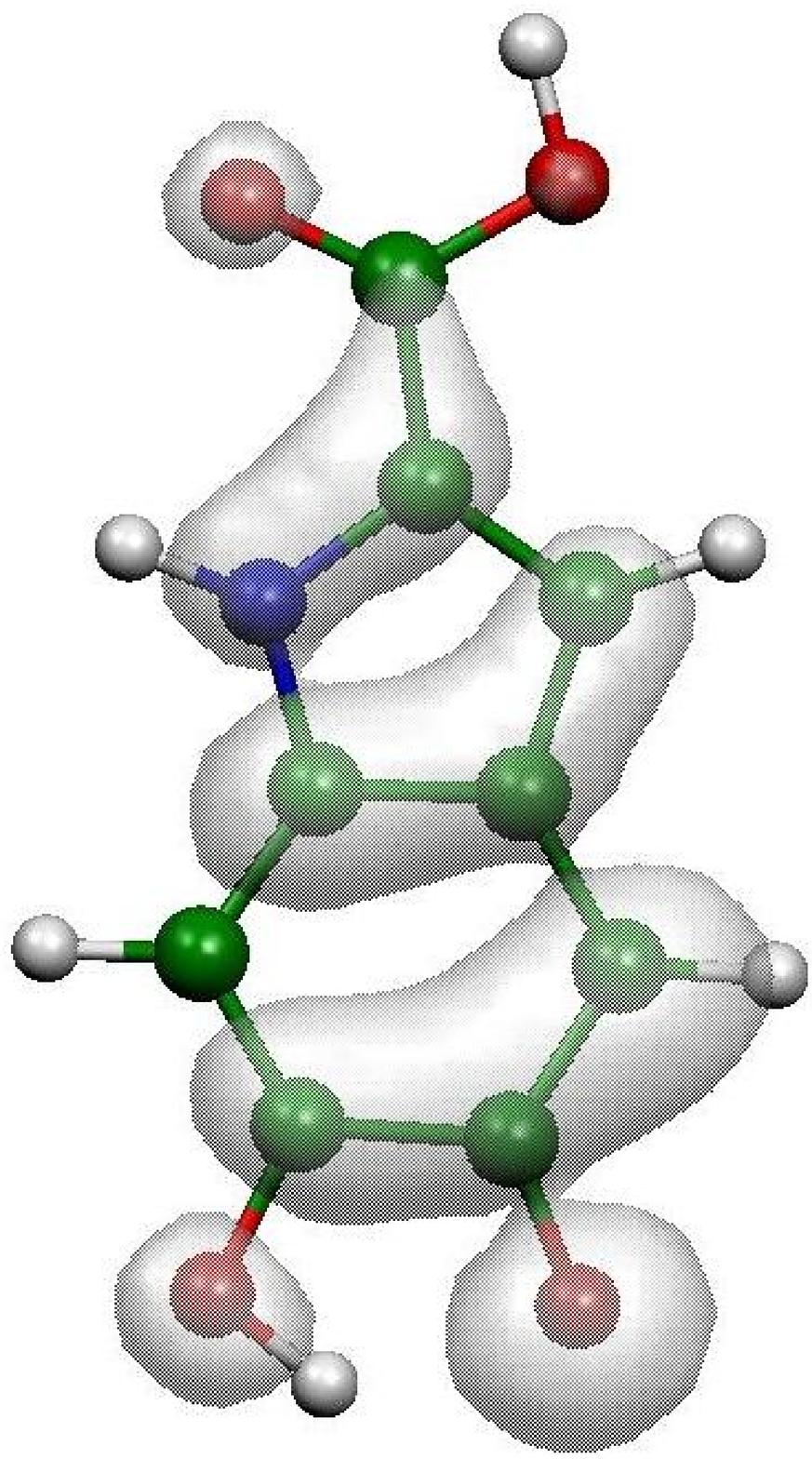, width=2.3cm, angle=90}
    \epsfig{figure=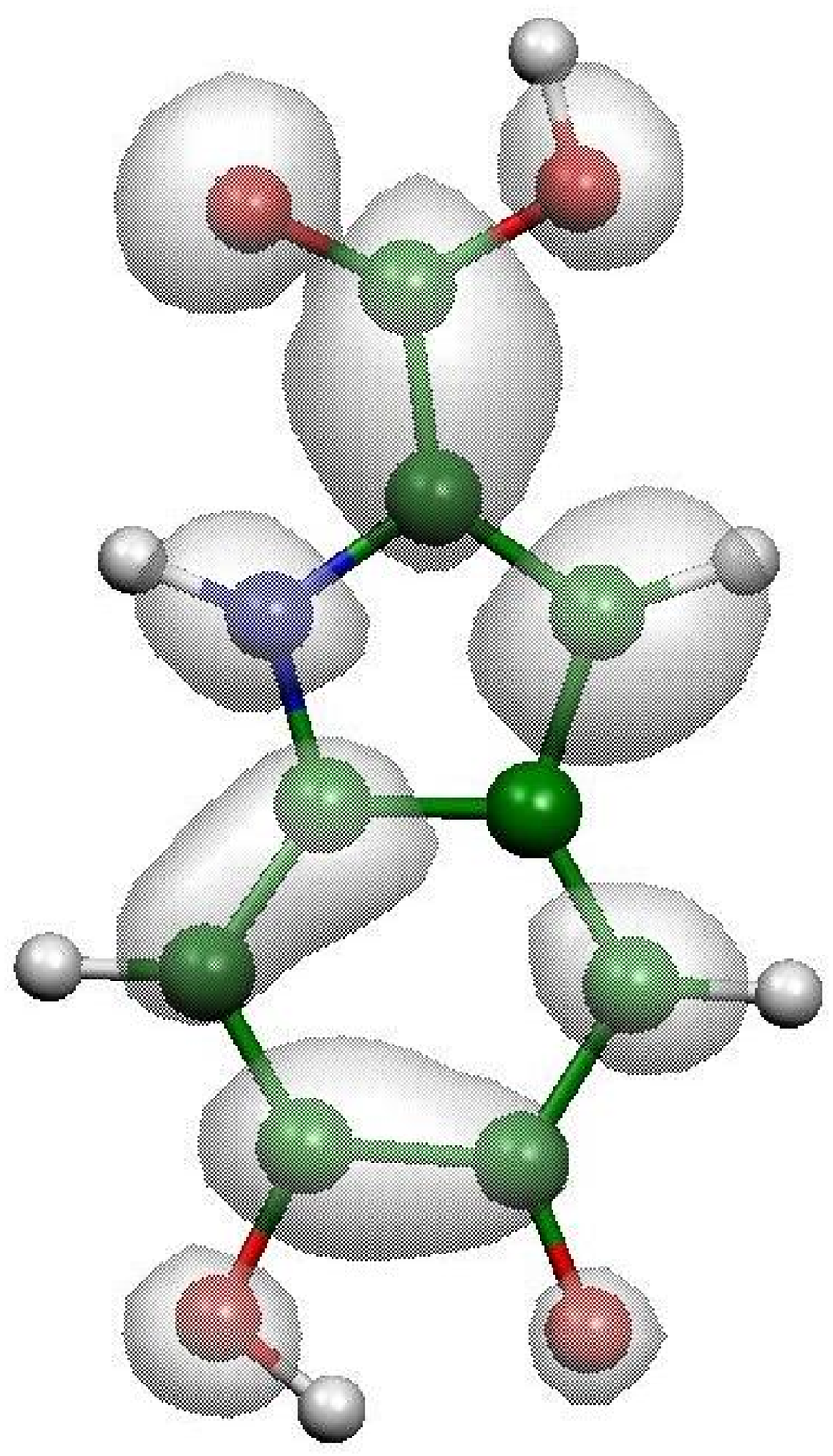, width=2.3cm, angle=90}
    \epsfig{figure=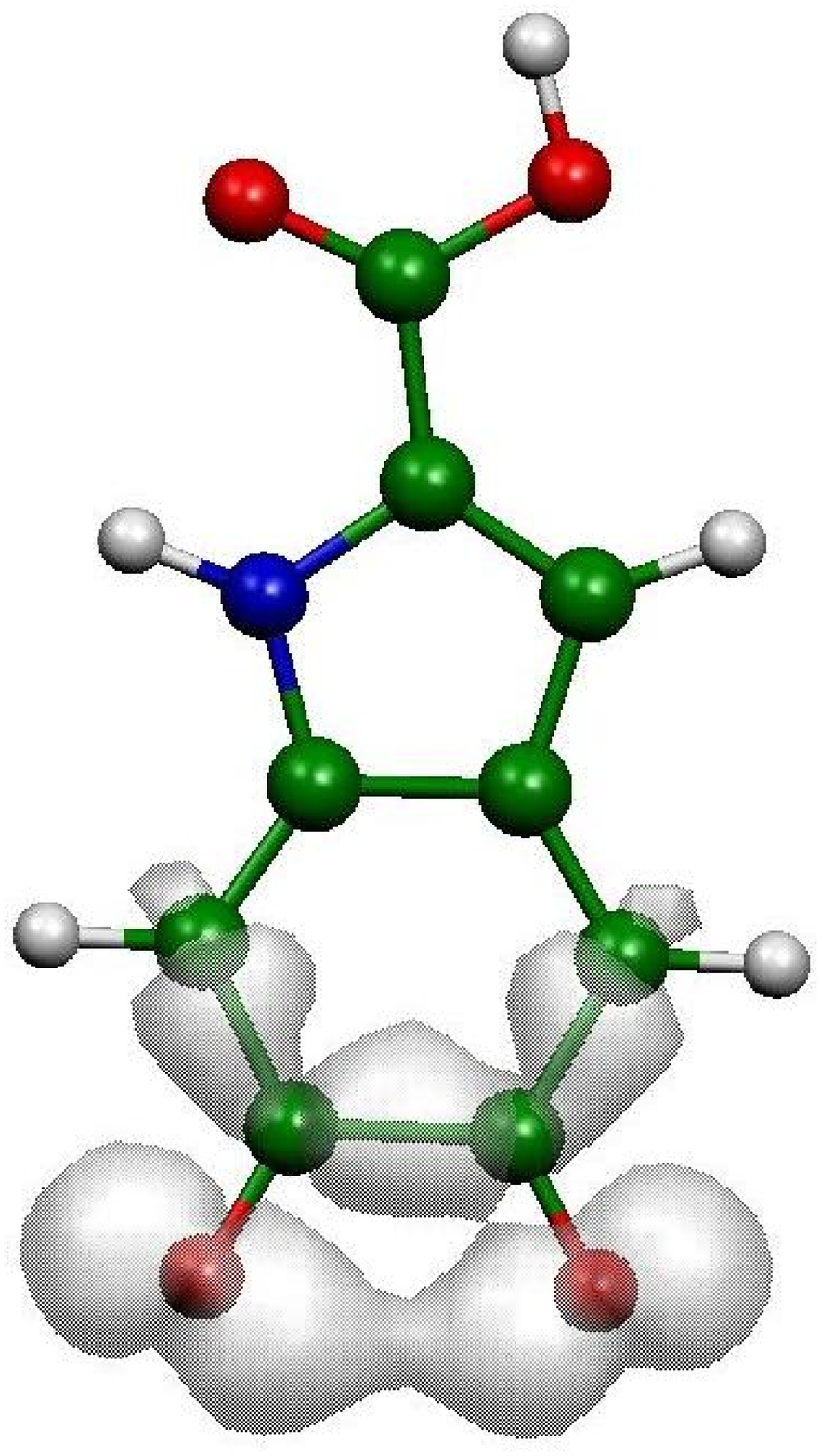, width=2.3cm, angle=90}
    \epsfig{figure=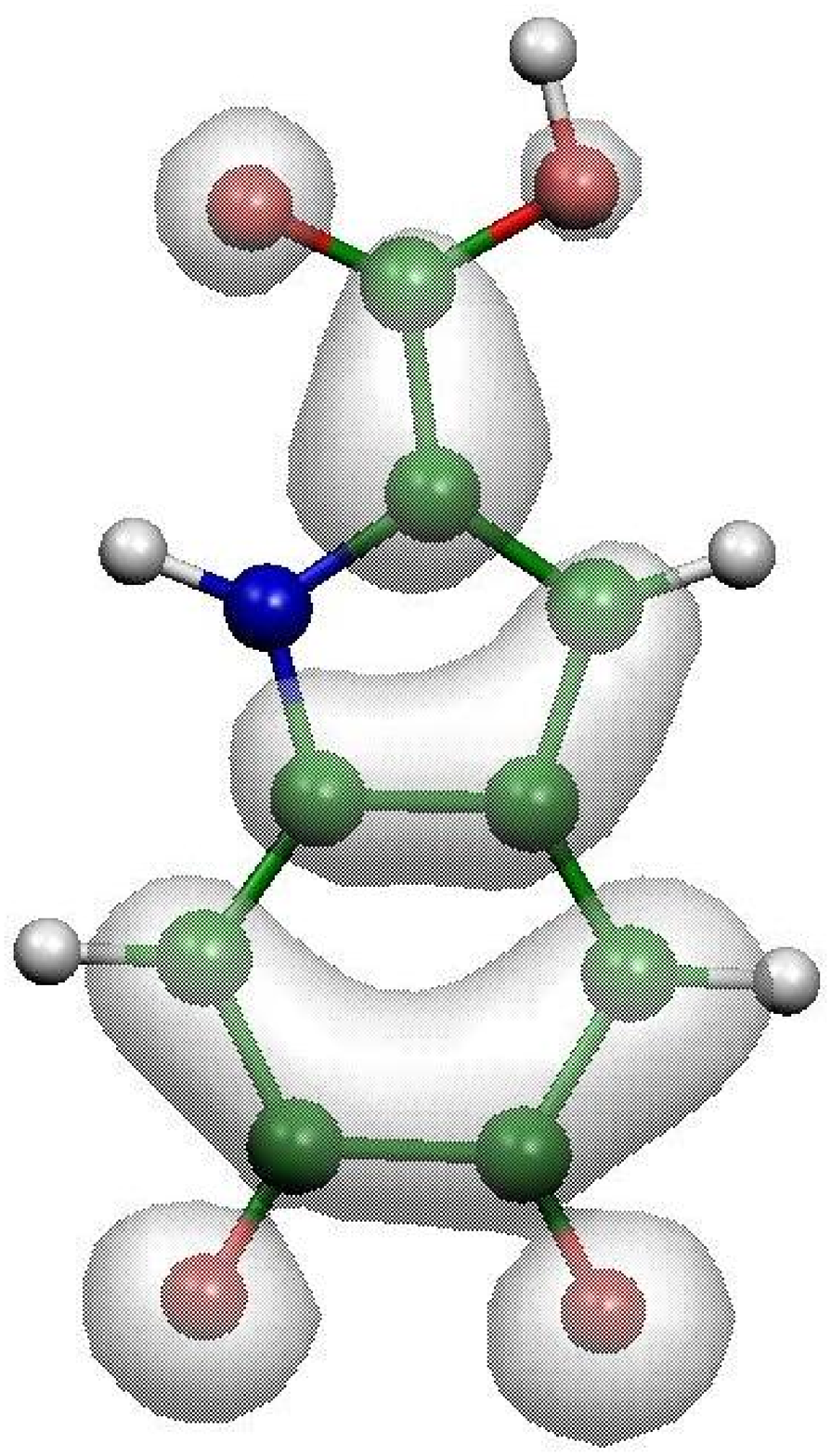, width=2.3cm, angle=90}
    \epsfig{figure=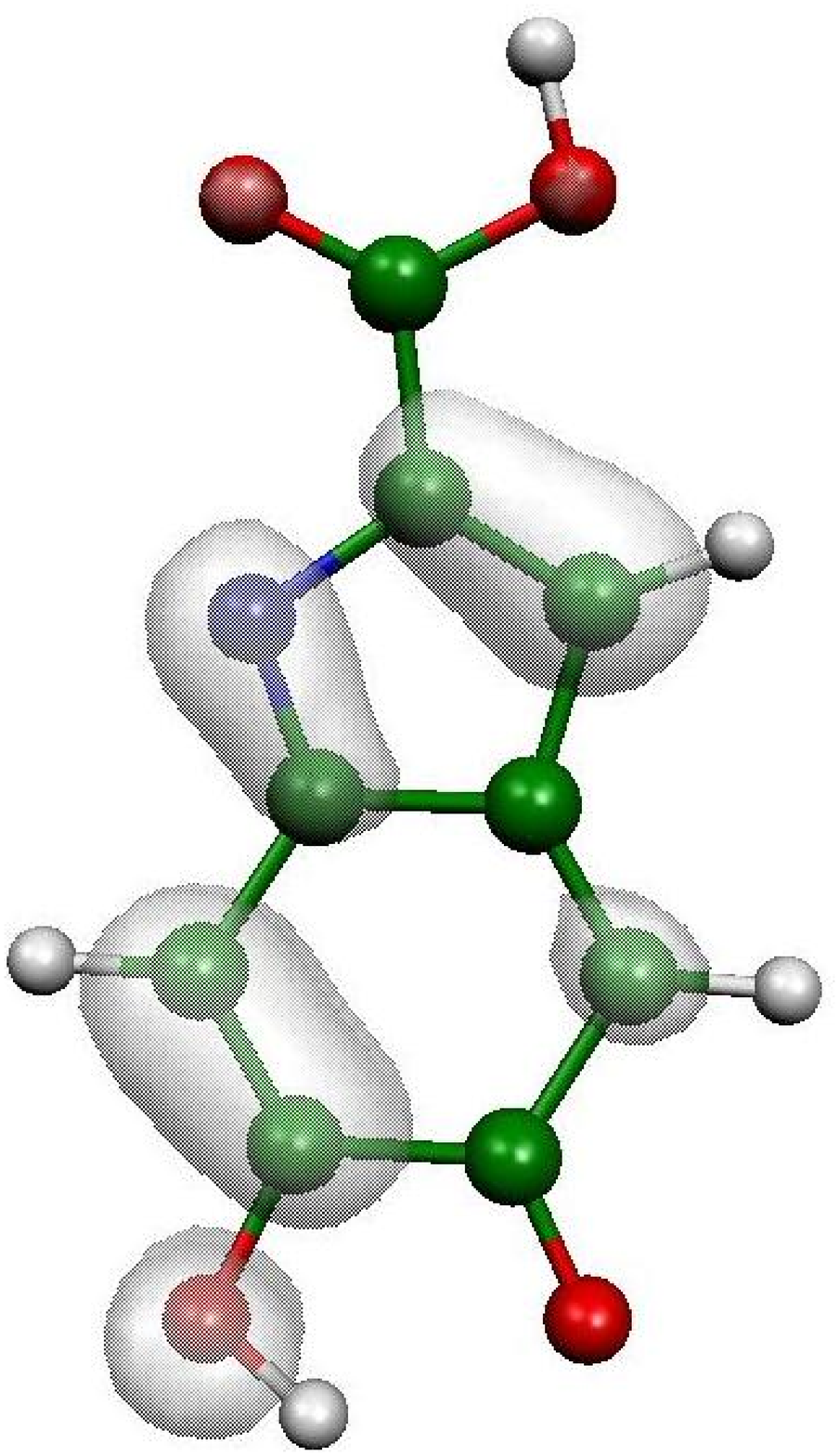, width=2.3cm, angle=90}
    \epsfig{figure=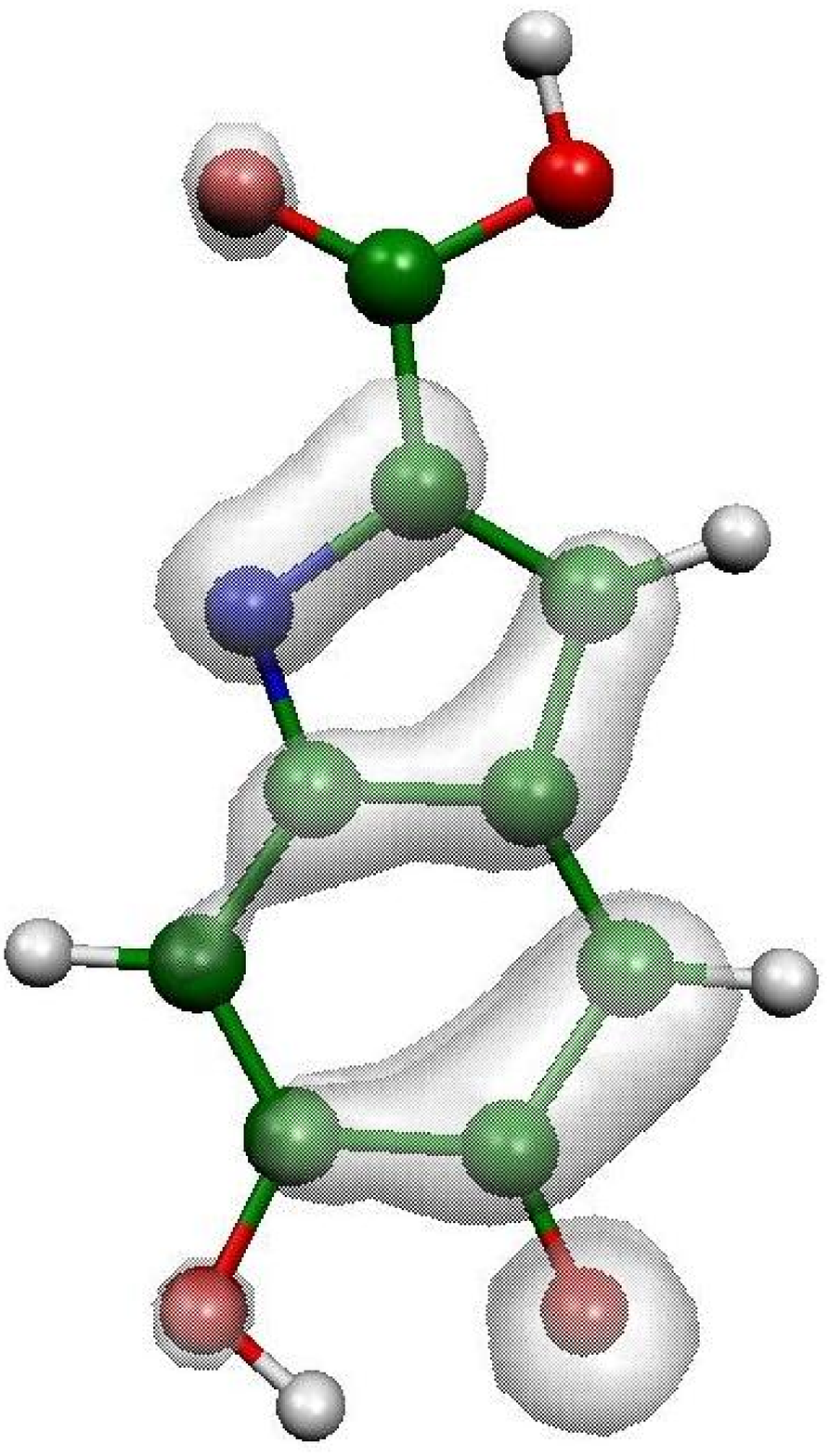, width=2.3cm, angle=90}
    \caption{(Colour online.) The electron density in the highest occupied molecular
        orbital (HOMO) (left) and the lowest unoccupied molecular
        orbital LUMO (right) of, from the top down, {\bf 1}, {\bf 2a}, {\bf 3a}
        and {\bf 3b}. 
        Although the electron
        density in the HOMO of  {\bf 1} is very similar to the electron density
        in the HOMO of DHI (reported in \cite{mel_mon}) there are significant differences between
        the electron densities in the LUMOs of DHICA and DHI near atom C2 (where the carboxylic
        group attaches). This is presumably related to the large
        difference ($\sim20\%$)
        in HOMO-LUMO gap calculated for the two compounds (c.f. table
        \ref{tab:HOMO-LUMO}). Both the HOMO and LUMO electron
        densities of {\bf 3a}
        and {\bf 3b}  are very similar to the electron density
        in the HOMO and LUMO of the equivalent
        reduced forms of DHI (the indolequinone and the semiquinone respectively, see \cite{mel_mon})
        This is presumably related to the similarity
        of HOMO-LUMO gap calculated for the carboxylated and uncarboxylated forms of the two compounds (c.f. table
        \ref{tab:HOMO-LUMO}).}
    \label{fig:density}
\end{figure}


One of the most important properties of melanin monomers is the
HOMO-LUMO gap. This is indicative of the optical spectrum as it
represents the fundamental optical absorption and H\"uckel model
calculations \cite{GC1} for DHI show that the HOMO-LUMO gap of a
single molecule correlates with the semiconducting gap of an
infinite homopolymer. The \lq\lq semiconductor" gap is likely to
be  a critical design parameter for molecularly engineered forms
of eumelanin which may be useful as functional materials in
electronic devices and sensors
\cite{Meredith&Riesz03,Paulschapter}.  Thus controlling the
HOMO-LUMO gap of monomers and thence of macromolecules may provide
a route to controlling the semiconducting gap of eumelanins
\cite{mel_mon}.

DFT is a theory of the ground state, therefore calculations
represent the energy gap between Kohn--Sham eigenvalues and not
the true HOMO-LUMO gap of the molecules. This is known as the band
gap problem \cite{Jones&Gunnarsson}. Additionally, it is accepted
that the PBE functional can significantly underestimate the
HOMO-LUMO gap. Therefore we have also employed the $\Delta$SCF
method \cite{Jones&Gunnarsson} to calculate the HOMO-LUMO gap. We
have previously shown that equivalent results for DHI reproduce
the trends found in time dependent DFT calculations
\cite{Ilichev}. In table \ref{tab:HOMO-LUMO} we compare the
HOMO-LUMO gap found from a simple interpretation of the Kohn--Sham
eigenvalues with those found by the $\Delta$SCF method. We also
reproduce the equivalent results for some of the uncarboxylated
forms from \cite{mel_mon}. From this it can be seen that for {\bf
3a} and {\bf 3b} carboxylation does not significantly change the
HOMO-LUMO gap. But, for {\bf 1}, which our results show is the
only significant component of DHICA samples (table
\ref{tab:abundances}), carboxylation causes a significant change
in the HOMO-LUMO gap. This indicates that, because natural
eumelanin contains both DHI and DHICA, a sample of eumelanin
contains a large range of HOMO-LUMO gaps, even before the effects
of the formation of macromolecules are considered.

The changes in the HOMO-LUMO gap in DHI are understood in terms of
the delocalisation of the wavefunction. Therefore in figure
\ref{fig:density} we plot the calculated electron density of {\bf
1}, {\bf 2a}, {\bf 3a} and {\bf 3b} respectively. Comparing these
plots with the equivalent results for the uncarboxylated forms
\cite{mel_mon} we see that for {\bf 3a} and {\bf 3b} the electron
densities in both the HOMOs and LUMOs is remarkably similar in
both the carboxylated (figure \ref{fig:density}) and
uncarboxylated forms. This is presumably the reason that the
calculated HOMO-LUMO gap is essentially independent of
carboxylation (table \ref{tab:HOMO-LUMO}). On the other hand while
the electron density of the HOMO of {\bf 1} is very similar to
that of its uncarboxylated form, there are significant differences
between the electron density of the LUMO and that of the
uncarboxylated form (figure \ref{fig:density}). This is,
presumably, associated with the large change in the HOMO-LUMO gap
caused by carboxylation (c.f. table \ref{tab:HOMO-LUMO}).

\section{Conclusions}

We have carried out first principles density functional
calculations for 5,6-dihydroxyindole-2-carboxylic acid (DHICA) and
several reduced forms. To the best of our knowledge these are the
first quantum chemical calculations for this system. These
molecules and similar molecules based on 5,6-dihydroxyindole (DHI)
are believed to be the basic building blocks of the eumelanins. We
used the $\Delta$SCF method to study the HOMO-LUMO gap. Comparing
our results for DHICA with previously published results for DHI we
found that that the addition of the carboxylic acid group has a
significant effect on the physical properties of the molecules and
in particular on the relative stability of the various redox forms
and the HOMO-LUMO gap. Based on the calculated relative
stabilities of the various redox forms in their various charge
states we find predict that there will be an extremely low
unpaired electron density in samples of pure DHICA (this is not
the case for DHI \cite{Bochenek,mel_mon}) and thus a negligible
signal should be seen in ESR experiments. This indicates that
experimental results cannot be straightforwardly extrapolated from
DHICA to DHI or \emph{vice versa}.

\section*{Acknowledgements}

This work was motivated by discussions with Paul Meredith. I thank
him and Linh Tran for sharing their results with me prior to
publication. I would also like to thank Tunna Baruah, Jan
Musfeldt, Mark Pederson and Jennifer Riesz for helpful
discussions. I thank Ross McKenzie for supporting this work and
for a critical reading of the manuscript. This work was funded by
the Australian Research Council. Calculations were performed on
the Australian Partnership for Advanced Computing (APAC) National
Facility under a grant from the Merit Allocation Scheeme.

\end{document}